\documentclass[1p]{elsarticle}
\usepackage{graphicx}
\usepackage{amssymb}
\usepackage{epsfig}
\usepackage{lineno}
\newcommand{\bm}[1]{\mbox{\boldmath$#1$}}
\def\bea{\begin{eqnarray}}
\def\eea{\end{eqnarray}}
\journal{Nuclear Instruments and Methods in Physics Research Section A}

\begin{document}

\begin{frontmatter}

\title{Coherent radiation of atoms and a channeling particle}
\author{V. Epp, M. A. Sosedova}
\address{Department of Theoretical Physics, Tomsk State Pedagogical University, Tomsk, 634061, Russia.}
\ead{epp@tspu.edu.ru, sosedova@tspu.edu.ru} 
\begin{abstract}
New mechanism of radiation emitted at channeling of a relativistic particle in a crystal is studied.  Superposition of coherent radiation of the atoms in a crystal lattice which are excited by a channeling particle and radiation of the channeling particle itself is considered. It is shown that coherent radiation of the chain of oscillating atoms forms a resonance peak on the background of radiation of the channeling particle.
\end{abstract}
\end{frontmatter}

\linenumbers

\section{Introduction}
Channeling of the charged particles was predicted by American scientists M.T. Robinson and O.S. Oen in 1961 \cite{Robin} and was discovered  by several groups of scientists in 1963 -- 1965.
Channeling of the accelerated particles was investigated in detail both theoretically and experimentally. The effect of channeling has served as the base for development of new experimental methods of research on the crystal structure.  It makes possible to measure distribution of electronic density in inter-atomic space of crystals and exact orientation of the crystal planes.
Channeling of ions is used recently for direct detections of  WIMP  (Weakly Interacting Massive Particle) -- the Dark Matter candidates \cite{Drob, Berna}.
 
At interaction between the channeling particle and a crystal the electromagnetic radiation is generated. In the framework of classical electrodynamics one can consider this radiation as radiation from different sources, such as radiation of the channeling particle, radiation of the electronic gas (wake trace) and radiation of the atoms excited by a channeling particle. Radiation of a channeling particle is investigated in details in many papers and books, see for example \cite{Lindhard, Baz}. Experimental study of radiation of wake fields can be found in \cite{Kil}. Less attention is paid to radiation of the atoms excited by a channeling particle. The last effect consists of two different parts -- radiation emitted at decay of exited state of atom, and radiation of vibrating atomic nucleus. When channeling, the charged particle transmits part of its transverse momentum to the surrounding atoms. This excites vibration of atomic nucleons together with the inner electrons. This phenomenon is similar to the vibrational excitation of molecules and causes respective radiation.  But vibrations of atoms in the channel are coherent, because they are exited by the same channeling particle. This can  lead to sufficient amplification of radiation intensity .
    
Assumption on radiation of this kind was expressed recently in \cite{Sosedova}. It was shown that radiation of an atomic chain is collimated in direction of the channeling particle velocity.  But proper radiation of the channeling particle was not taken into account. Actually, the field of radiation should be calculated as a superposition of fields generated both by channeling particle and excited atoms. 

In this paper we study the field of vibrating atoms in some more details and calculate angular distribution of radiation resulting from superposition of fields produced by  channeling particle and by atoms. In section \ref{sec2} we discuss some features of radiation field of atomic chain which were not studied in  \cite{Sosedova}. Next section presents calculation of the net field  produced by channeling particle and excited atoms. We discuss the ratio of radiation intensity caused by vibrating atoms and by channeling particle.  Angular distribution of radiation is plotted.

\section{Field of the atomic chains exited by a channeling particle}\label{sec2}

\begin{figure}[htbp]\center
\includegraphics[width=2in]{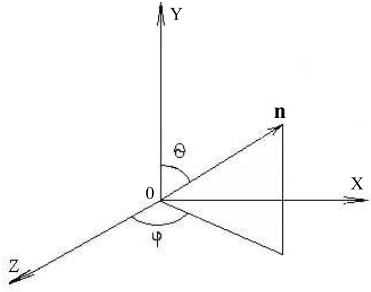}
\caption{The coordinate system.}
\label{fig1}
\end{figure}

Electric field $E_{\rm a}(t)$ of vibrating atoms in far zone is given by equation (30) in  \cite{Sosedova}. We rewrite this equation in a complex form
\bea\label{eott2}
E_{\rm a}(t)=\frac{A_0\nu^2}{2k}\left\{\frac{e^{i\nu t}-e^{-i\Omega 't}}{\nu+\Omega'}+\frac{e^{i\nu t}-e^{i\Omega 't}}{\nu-\Omega'}\right\},
\eea
for $t\leq Lk$ and
\bea\label{eott3}
E_{\rm a}(t)=\frac{A_0\nu^2e^{i\nu t}}{2k}\left\{\frac{1-e^{-ikL(\nu+\Omega')}}{\nu+\Omega'}+\frac{1-e^{-ikL(\nu-\Omega')}}{\nu-\Omega'}\right\},
\eea
for $t> Lk$. Here 
\begin{equation} \label{amplit}
A_0 =\frac {2eq^2K\sin\theta} {mVD^2\omega bRc^2},
\quad k=\frac {1}{V} (1-\beta\sin\theta\sin\varphi),\quad \nu=\omega+i\alpha,
\end{equation}
$V$   is the mean velocity of the channeling particle, $\beta=V/c$, $c$ is the speed of light, $e$ is the charge of the channeling particle, $m$ is the mass of atom,  $2D$ is the width of the channel, $b$ is the spacing between atoms in an atomic chain and $q$ is an effective charge of a nucleus of atom which is screened by the interior electrons. Angles $\theta$ and $\varphi$ of spherical coordinate system are depicted in Fig. \ref{fig1}  $K$ and $\Omega$ are the amplitude and frequency of channeling particle oscillations which are bound to $XOY$ plane. The mean velocity of the particle is directed along the $X$-axis, the particle enters the crystal at $x=0$ and leaves it at $x=L$.
An exited atom of the crystal lattice is oscillating along the $Y$-axis with frequency $\omega$ and amplitude  decreasing in time as ${\rm exp}(-\alpha t)$, where $\alpha$ is the attenuation factor.

 Further we consider only this part of radiation. Taking the real part of equation (\ref{eott2}) we obtain
 \bea\label{eott5}
E_{\rm a}(t) &=& -\frac{A_0\omega}{k\left[(\alpha^2+\omega^2-\Omega'^2)^2+4\alpha^2\Omega'^2\right]}\left\{2\alpha\Omega'^3\sin\Omega't\right.\nonumber\\
&+&\left.\cos\Omega't\left[\left(\alpha^2 +\omega^2\right)^2+\Omega'^2(3\alpha^2-\omega^2)\right]\right\}.
\eea
This field has some features typical for the field of relativistic charged particle though the radiation is caused by non-relativistic vibrating atoms. First, the multiple $1-\beta  \sin\theta\sin\varphi$ in denominator of $A_0$ shows that main part of radiation is concentrated around the direction of the particle velocity $\bm \beta$.  Second, the frequency of radiation $\Omega'$ is just the same as the frequency of the relativistic channeling particle.
Coherence of radiation is expressed by factor $b$ in denominator of $A_0$. It shows that intensity of radiation is proportional to the square of  linear charge density $q/b$. Denominator of this expression is of typical resonant character. Indeed, if attenuation of atomic vibrations vanishes ($\alpha\to0$) and $\Omega'=\omega$, the electric field defined by the above equation tends to infinity. 
\section{Superposition with the field of channeling particle}
Radiation of the vibrating atoms is emitted at the same frequency $\Omega'$ as the radiation of the channeling particle. And in case of relativistic particle it is concentrated in the same narrow cone around the direction of the particle velocity. Hence, the fields of both sources are correlated   and the resulting field must be calculated as superposition of the fields.

Let us calculate the  radiation field of the channeling particle. Electric field $\bm E_c$  of a moving charge $e$  is given by equation \cite{Landau}:
\begin{equation}\label{Pole1}
\bm{E}_{\rm c}=\displaystyle\frac{e}{c^2R(1-\bm\beta \bm{n})^3}\left[\bm{n}\left[\left(\bm{n}-\bm\beta \right)\dot{\bm{v}}\right]\right],
\end{equation}
where $\bm v$ is the velocity of charged particle and $\bm \beta=\bm v/c$. Next, we assume that the transversal component $v_\perp$ of the particle velocity is small enough to satisfy inequality $v_\perp\ll c\sqrt{1-\beta^2}$. It is shown in \cite{Landau} that in this case we can consider  $\bm v$ as being constant, keeping as time dependent only $\dot{\bm v}$. Hence, we put  $\bm v= (V,0,0)$. 

Substituting the low of motion $x(t)=V t$, $y(t)=K\cos\Omega t$, $z(t)=0$ into (\ref{Pole1}) we get the electric field components in spherical coordinates defined by Fig. \ref{fig1}
\begin{equation}\label{Komponents}
{E_{{\rm c}\theta}=\displaystyle\frac{eK\Omega^2(\beta\sin\varphi-\sin\theta)\cos\Omega' t}{c^2R(1-\beta\sin\theta\sin\varphi)^3}, \,
E_{{\rm c}\varphi}=\displaystyle\frac{eK\Omega^2\beta\cos\theta\cos\varphi\cos\Omega' t}{c^2R(1-\beta\sin\theta\sin\varphi)^3}.}
\end{equation}

The net radiation field $\bm E$ is a vector sum of the fields of atomic chains  (\ref{eott5}) and of the channeling particle (\ref{Komponents}).
The electric field $\bm E_{\rm a}$ is polarized in a plane which includes axis $OY$.
In order to compute the intensity of radiation we average over time the square of the resulting  field. This yields
\begin{eqnarray}\label{srednee1}
&&\overline{E_\theta^2}=E_0^2\left\{\frac{\left(\beta\sin\varphi-\sin\theta\right)^2}{\left(1-\beta\sin\theta\sin\varphi\right)^4}\right.\nonumber\\
&&-2\eta\gamma^2\frac{\sin\theta\left(\beta\sin\varphi-\sin\theta\right)\left[\left(\alpha^2+\omega^2\right)^2+ {\Omega'}^2\left(3\alpha^2-\omega^2\right)\right]}{\left(1-\beta\sin\theta\sin\varphi\right)^2\left[(\alpha^2+\omega^2-\Omega'^2)^2+4\alpha^2\Omega'^2\right]} \nonumber\\
&&+\left. \eta^2\gamma^4\frac{\sin^2\theta\left[\left(\alpha^2+\omega^2\right)^2+4\alpha^2{\Omega'}^2\right]}{\left[(\alpha^2+\omega^2-\Omega'^2)^2+4\alpha^2\Omega'^2\right]}\right\},\\
\label{srednee2}
&&\overline{E_\varphi^2}=E_0^2\frac{{\beta}^2\cos^2\theta\cos^2\varphi}{(1-\beta\sin\theta\sin\varphi)^4},
\end{eqnarray}
where
\begin{eqnarray}
E_0=\displaystyle\frac{e\Omega^2K}{\sqrt{2}\,c^2R(1-\beta n_x)}, \nonumber
\end{eqnarray}
\begin{eqnarray}
\eta=\frac{2{q}^2}{m\gamma^2\Omega^2bD^2}, \quad
\gamma=\frac{1}{\sqrt{1-\beta^2}}. \nonumber
\end{eqnarray}
Factor $\eta$ shows the relation between the intensity of radiation generated by atoms and radiation emitted by the channiling particle.

For an ultra-relativistic particle we have only to consider small angles around the velocity direction and can make the usual approximation, leading to reduced angles
\[\psi_y=\gamma\left(\frac\pi 2-\theta\right),\quad \psi_z=\gamma \left(\frac \pi 2 -\varphi\right),\quad \psi^2=\psi_y^2+\psi_z^2,\]
In this approximation (\ref{srednee1}) and (\ref{srednee2}) take the form:
\begin{eqnarray}\label{aproxtheta}
&&\overline{E_\theta^2}=E_0^2\gamma^4\left\{4\frac{\left(1-{\psi_y}^2+{\psi_z}^2\right)^2}{\left(1+\psi^2\right)^4}\right.\nonumber\\
&&+4\eta\frac{\left(1-{\psi_y}^2+{\psi_z}^2\right)^2\left[\left(\widetilde{\alpha}^2+1\right)^2+\widetilde{\Omega}'^2\left(3\widetilde{\alpha}^2-1\right)\right]}{\left(1+\psi^2\right)^2\left[(\widetilde{\alpha}^2+\omega^2-\widetilde{\Omega}'^2)^2+4\widetilde{\alpha}^2\widetilde{\Omega}'^2\right]}\nonumber\\
&&+\left.\eta^2\frac{\left(\widetilde{\alpha}^2+1\right)^2+4\widetilde{\alpha}^2\widetilde{\Omega}'^2} {\left[(\widetilde{\alpha}^2+\omega^2-\widetilde{\Omega}'^2)^2+4\widetilde{\alpha}^2\widetilde{\Omega}'^2\right]}\right\},\\
\label{aproxphi}
&&\displaystyle \overline{E_\varphi^2}=32E_0^2\,\frac{\gamma^4{\psi_y}^2{\psi_z}^2}{(1+\psi^2)^4},
\end{eqnarray}
where
\begin{equation}\label{for}
\widetilde{\alpha}=\alpha/\omega, \quad
\widetilde{\Omega'}=\Omega'/\omega.
\end{equation}
 Let us estimate the value of $\eta$. The frequency $\Omega$ depends on the continuous interaction potential of atomic strings with the channeled particle and is of order $\Omega\approx c\psi/D$,  where $\psi$ is the Lindhard angle
\begin{equation}\label{angle}
\psi=\sqrt{\frac{eq}{{\cal E}b}},
\end{equation}
${\cal E}=m_ec^2\gamma$ is the energy of the channeling particle, $m_e$ is its mass. Substituting all this into expression for $\eta$ we get
\[\eta\approx\frac{qm_e}{em\gamma}.\]
If the channeling particle is a proton, the charge of nucleus is $q=Ze$ and mass of nucleus is $m=2Ze$, then $\eta\approx\gamma^{-1}$.
 But we have to take into account the resonant character of the last term in equation (\ref{aproxtheta}). Substituting $\widetilde{\Omega}'=1$ into (\ref{aproxtheta}) we see that  the ratio of last term to the first one is of order $\eta^2\alpha^{-2}\sim[m_e/(m\gamma\alpha)]^2$.

Hence,  in case of heavy channeling particle and low attenuation the radiation intensity of the atomic string can be comparable with  radiation intensity of the channeling particle. The resonance condition
\begin{equation}
\widetilde{\Omega'}=2\gamma^2\frac{\Omega}{\omega(1+\psi^2)}=1
\end{equation} 
can be fulfilled only if $2\gamma^2\Omega/\omega>1$.

As an example we have plotted angular distribution of radiation intensity 
\[I_i(\psi_y, \psi_z)=\frac{c}{4\pi}R^2 E_i^2, \quad i=\theta,\varphi\]
using formula for the field components (\ref{aproxtheta}) and (\ref{aproxphi}). We put $\eta=10^{-2}$, $\gamma=100$, $\widetilde\alpha=10^{-2}$.

\begin{figure}[ht]
\center{\includegraphics[width=2in]{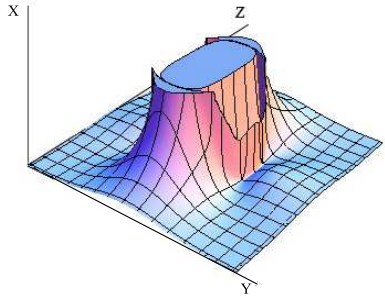}}
\caption{Polar component of radiation intensity $I_\theta$.}
\label{fig31}
\end{figure}

\begin{figure}
\center{\includegraphics[width=2in]{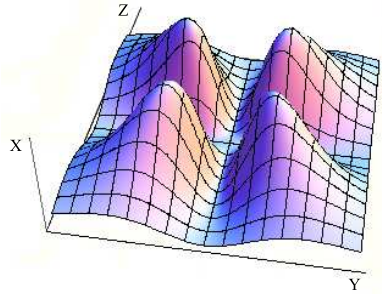}}
\caption{Azimuthal component of radiation intensity $I_\varphi$.}
\label{fig32}
\end{figure}

Polarization components of radiation are  shown in the Figs \ref{fig31} and
 \ref{fig32}. Axes $Y$ and $Z$ correspond to $\psi_y$ and $\psi_z$ respectively. All figures are plotted for interval $-1<\psi_y<1$, $-1<\psi_z<1$.
\hfill

\begin{figure}[ht]\center
\includegraphics[width=2.7in]{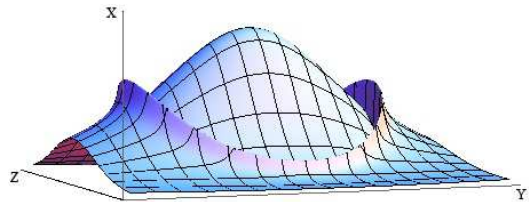}
\caption{Sum of polarization components.}
\label{fig3}
\end{figure}

We see that the main part of radiation intensity is the radiation of the channeling particle which in our case is represented by radiation of an harmonic oscillator moving with constant velocity. Radiation of this source  is investigated in details in many papers (see for example \cite{Bordovitsyn} and references therein). Radiation  emitted by the vibrating atoms is represented as a sharp peak at definite angle $\psi$. The angular distribution of total radiation intensity $I=I_\theta+I_\varphi$  is depicted in Fig. \ref{fig3}.

\section{Conclusion}

At first glance, radiation of atoms exited by a channeling particle and 
vibrating in the points of crystal lattice is negligible small because of
small amplitude and small frequency of vibrations, which is a
consequence of relatively great mass of atoms. But if we take into account
that great number of atoms are involved in coherent oscillations by the
channeling particle, we get sufficient amplification of the intensity of
radiation.
Derived expressions show that this kind of radiation can  be comparable in
magnitude with radiation of the channeling particle itself.
It is shown that the vibrating atoms produce radiation field which
results in a sharp peak at certain directions on the background of radiation
of the channeling particle. This effect can be observed at channeling of
heavy relativistic particles such as protons or ions.
Very simplified model of crystal lattice was used just to show the essence
of the phenomenon. Many details have not been taken into account.

For example, this paper has not reviewed thermal vibrations of atoms.
These can be taken into account by adding random oscillations to the regular harmonic oscillations. This will cause additional background in intensity of
radiation. Evidently, the  momentum inhomogeneity in a beam of the
channeling  particles will broaden the resonant peak of radiation shown 
in Figs \ref{fig31} and \ref{fig3}.

\section*{Acknowledgement}
This research has been supported by the grant for LRSS, project No 224.2012.2.

\end{document}